\documentstyle [twocolumn,epsf]{mn}
\newcommand{\mincir}{\raise
-3.truept\hbox{\rlap{\hbox{$\sim$}}\raise4.truept\hbox{$<$}\ }}
\newcommand{\magcir}{\raise
-3.truept\hbox{\rlap{\hbox{$\sim$}}\raise4.truept\hbox{$>$}\ }}
\newcommand{\minmag}{\raise
-3.truept\hbox{\rlap{\hbox{$<$}}\raise5.truept\hbox{$<$}\ }}
\newcommand{\be}{\begin{equation}}
\newcommand{\ee}{\end{equation}}
\newcommand{\ba}{\begin{eqnarray}}
\newcommand{\ea}{\end{eqnarray}}
\newcommand{\brr}{\begin{array}}
\newcommand{\err}{\end{array}}
\newcommand{\bc}{\begin{center}}
\newcommand{\ec}{\end{center}}

\renewcommand{\baselinestretch}{1.0}

\title[The growth index of matter perturbations and modified gravity]
{The growth index of matter perturbations and modified gravity}
\author[Spyros Basilakos \& Athina Pouri]{Spyros Basilakos$^1$ \& Athina
Pouri$^{1,2}$\\
\vspace{0.1cm}
$^1$ Academy of Athens, Research Center for Astronomy \& Applied
  Mathematics, Soranou Efessiou 4, 11-527, Athens, Greece\\
$^2$ Faculty of Physics, Department of Astrophysics - Astronomy -
Mechanics University of Athens, Panepistemiopolis, Athens 157 83
}

\begin{document}

\maketitle

\begin{abstract}
We place tight constraints on the growth index $\gamma$
by using the recent growth history results of
2dFGRS, SDSS-LRG, VIMOS-VLT deep Survey (VVDS)
and {\em WiggleZ} datasets. In particular,
we investigate several parametrizations of the growth
index $\gamma(z)$, by comparing their cosmological
evolution using observational growth rate data
at different redshifts.
Utilizing a standard likelihood analysis we find that the use of
the combined growth data provided by the 2dFGRS, SDSS-LRG, VVDS
and {\em WiggleZ} galaxy surveys,
puts the most stringent constraints on the value of the
growth index. As an example, assuming a constant growth index
we obtain that $\gamma=0.602\pm 0.055$ for the
concordance $\Lambda$CDM expansion model.
Concerning the Dvali-Gabadadze-Porrati gravity model,
we find $\gamma=0.503\pm 0.06$
which is lower, and almost $3\sigma$ away, from
the theoretically predicted value of $\gamma_{DGP}\simeq 11/16$.
Finally, based on a time varying growth index
we also confirm that the combined growth data disfavor the DGP gravity.

\end{abstract}

\vspace{1.0cm}

\section{Introduction}
Recent studies in observational cosmology, using all
the available high quality cosmological data (Type Ia
supernovae, cosmic microwave background, baryonic acoustic
oscillations, etc), converge to
an emerging ``standard model''. This cosmological model is spatially flat
with a cosmic dark sector usually formed by cold dark matter and some
sort of dark energy, associated with large negative pressure, in order to
explain the observed accelerating expansion of the Universe
(cf. Tegmark et al. 2004; Spergel et al. 2007; Davis et al. 2007;
 Kowalski et al. 2008; Hicken et al. 2009;
Komatsu et al. 2009; Hinshaw et al. 2009; Lima \& Alcaniz 2000; Jesus \& Cunha 2009; Basilakos \& Plionis 2010; Komatsu et al. 2011
and references therein).
Despite the mounting
observational evidence on the existence of the dark energy
component in the universe, its nature and fundamental origin remains
an intriguing enigma challenging the very foundations of theoretical
physics.
Indeed, during the
last decade there has been an intense  theoretical debate among cosmologists
regarding the nature of the exotic ``dark energy''.
The absence of a fundamental physical theory, concerning
the mechanism inducing the cosmic acceleration, has opened a window
to a plethora of alternative cosmological scenarios.
Most are based either on the existence of new fields in nature (dark
energy) or in some modification of Einstein's general relativity,
with the present accelerating stage appearing as a sort of
geometric effect (for reviews see Copeland, Sami \& Tsujikawa 2006;
Caldwell \& Kamionkowski 2009; Amendola \& Tsujikawa 2010 and
references therein).

In order to test the validity of general relativity on cosmological scales,
it has been proposed that measuring
the so called growth index, $\gamma$, could provide an efficient
way to discriminate between scalar field dark energy
(hereafter DE) models which admit to general relativity and
modified gravity models (cf. Ferreira \& Skordis 2010 and references therein).
Linder \& Cahn (2007) have shown that
there is only a weak dependence of $\gamma$ on
the equation of state parameter $w(z)$, implying
that one can separate the background expansion history, $H(z)$,
constrained by a large body of cosmological data (SNIa, BAO, CMB etc),
from the fluctuation growth history, given by $\gamma$.
In this framework, it was theoretically found that for those
DE models which adhere to
general relativity the growth index $\gamma$ is
well approximated by
$\gamma \simeq \frac{3(w-1)}{6w-5}$
(see Silveira \& Waga 1994; Wang \& Steinhardt 1998; Linder 2004;  
Linder \& Cahn 2007; Nesseris \& Perivolaropoulos 2008; 
Lee \& Kin-Wang 2010), which
reduces to $\gamma_{\Lambda} \simeq 6/11$ for the
traditional $\Lambda$CDM cosmology $w(z)=-1$.
On the other hand, in the case of the
braneworld model of Dvali, Gabadadze and Porrati (2000; hereafter DGP)
the growth index becomes
$\gamma_{DGP} \simeq 11/16$ (see also Linder 2004; Linder et al 2007; 
Gong 2008; Wei 2008; Fu, Wu \& Hu 2009), while for
the $f(R)$ gravity models we have $\gamma \simeq 0.41-0.21z$
for $\Omega_{m0}=0.27$ (Gannouji, Moraes \& Polarski 2009; 
Tsujikawa et al. 2009; Motohashi, Starobinsky \& Yokoyama 2010).
Indirect methods
to determine $\gamma$ have also been proposed (mostly using a
constant $\gamma$),
based either on the observed growth rate of clustering
(Nesseris \& Perivolaropoulos 2008; Guzzo et al. 2008; Di Porto \& Amendola 2008;
Gong 2008; Dosset et al. 2010;
Samushia, Percival \& Racanelli 2012; Hudson \& Turnbull 2012)
providing a wide range of $\gamma$ values
$\gamma=(0.58-0.67)^{+0.11\; +0.20}_{-0.11 \; -0.17}$,
or on massive galaxy clusters Vikhlinin et al. (2009) and
Rapetti et al. (2010) with the
latter study providing $\gamma=0.42^{+0.20}_{-0.16}$,
or even on the weak gravitational lensing Daniel et al. (2010).
Gaztanaga et al. (2012) performed
a cross-correlation analysis between
probes of weak gravitational lensing and
redshift space distortions and found no evidence for deviations
from general relativity.
With the next generation of surveys, based on {\em Euclid} and
  {\em BigBOSS},
we will be able to put strong constraints on $\gamma$
(see for example Linder 2011; Belloso, Garcia-Bellido \& Sapone 2011;
Di Porto, Amendola \& Branchini 2012 
and references therein) and thus
to test the validity of general relativity
on extragalactic scales.

The scope of the present study is along the same lines,
ie., to place constraints on the growth index
using a single cosmologically relevant experiment, ie., that of the
recently derived growth data of the 2dFGRS, SDSS-LRG, VVDS
and {\em WiggleZ} galaxy surveys.
Note that for the background we use two reference expansion models
namely flat $\Lambda$CDM and DGP respectively.
The interesting aspect of the latter scenarios is that the
corresponding functional forms of the Hubble parameters
are affected only by one free parameter, that of
the dimensionless matter density at the present time
$\Omega_{m0}$.
The structure of the article is as follows.
In section 2, we briefly discuss the background
cosmological equations.
The theoretical elements of
the growth index are presented in section 3 in which
we extend the original Polarski \& Gannouji (2008) method
for a large family of $\gamma(z)$ parametrizations.
In section 4 we briefly discuss the growth data.
In section~5, we perform a likelihood analysis in order
to constrain the growth index model
free parameters. Finally, the main conclusions are
summarized in section 6.

\section{The background evolution}
For homogeneous and isotropic flat cosmologies, driven by
non relativistic matter and an
exotic fluid (DE models) with equation of state
(hereafter EoS), $p_{DE}=w(a)\rho_{DE}$, the
first Friedmann equation can be written as:
\begin{equation}
\frac{H^{2}(a)}{H_{0}^{2}}\equiv
E^{2}(a)=
\Omega_{m0}a^{-3}+\Omega_{DE,0}{\rm
e}^{3\int^{1}_{a} d{\rm lny}[1+w(y)]},  \label{nfe1}
\end{equation}
where $E(a)$ is the normalized Hubble flow,
$a(z)=1/(1+z)$ is the scale factor of the universe,
$w(a)$ is the EoS parameter, $\Omega_{m0}$ is the
dimensionless matter density at the present time and
$\Omega_{DE,0}=1-\Omega_{m0}$ denotes the DE density parameter.
Using the Friedmann equations,
it is straightforward to
write the EoS parameter in terms of $E(a)=H(a)/H_{0}$ (Saini et al. 2000;
Huterer \& Turner 2001)
\begin{equation}
\label{eos22}
w(a)=\frac{-1-\frac{2}{3}a\frac{{d\rm lnE}}{da}}
{1-\Omega_{m}(a)}
\end{equation}
where
\be
\label{ddomm}
\Omega_{m}(a)=\frac{\Omega_{m0}a^{-3}}{E^{2}(a)} \;.
\ee
Differentiating the latter and taking into account
eq.~(\ref{eos22}) we obtain
\be
\label{domm}
\frac{d\Omega_{m}}{da}=
\frac{3}{a}w(a)\Omega_{m}(a)\left[1-\Omega_{m}(a)\right]\;.
\ee
Since the exact nature of the DE is unknown, the
above DE EoS parameter includes our ignorance regarding the
physical mechanism powering the late time cosmic acceleration.
It is also worth noticing that the
concordance $\Lambda$CDM cosmology is described by a DE model with
$w(a)=-1$.

Interestingly, the above method can be generalized
to the context
of modified gravity. 
Indeed, instead of using the exact Hubble flow through a modification of the
Friedmann equation one may consider an equivalent Hubble flow
somewhat mimicking  eq. (\ref{nfe1}). The ingredient here is that
the accelerating expansion can be attributed to a kind of
``geometrical'' DE contribution. Now, due to the fact that the matter
density (baryonic+dark) cannot accelerate the cosmic expansion, it is fair to
utilize the following parametrization (Linder \& Jenkins 2003; Linder 2004):
\begin{equation}
E^{2}(a)=\frac{H^{2}(a)}{H_{0}^{2}}= \Omega_{m0}a^{-3}+\Delta H^{2}.
\label{nfe2}
\end{equation}
It becomes clear that any modification to the Friedmann equation of general
relativity is included in the last term of the above expression.
Now using eqs. (\ref{eos22}) and (\ref{nfe2}) one can derive
the effective (``geometrical'') dark energy EoS
parameter
\begin{equation}
\label{eos222}
w(a)=-1-\frac{1}{3}\;\frac{d{\rm ln}\Delta
H^{2}}{d{\rm ln}a}.
\end{equation}

In the context of a flat DGP cosmological model the
''accelerated'' expansion of the universe can be explained by a
modification of the gravitational interaction in which gravity
itself becomes weak at very large distances (close to the Hubble
scale) due to the fact that our four dimensional brane survives into
an extra dimensional manifold (see Deffayet, Dvali \& Cabadadze 2002
and references therein). An interesting feature
of this pattern is that the
corresponding functional form of the normalized Hubble function as given
by  eq. (\ref{nfe2}), contains only one free parameter,
$\Omega_{m0}$. The quantity $\Delta H^{2}$ is given by
\be
\Delta H^{2}=2\Omega_{bw}+2\sqrt{\Omega_{bw}}
\sqrt{\Omega_{m0}a^{-3}+\Omega_{bw}}
\ee
where
$\Omega_{bw}=(1-\Omega_{m0})^{2}/4$. From eq.(\ref{eos222}), it is
readily checked that the geometrical (effective)
DE equation of state parameter reduces to \be
w(a)=-\frac{1}{1+\Omega_{m}(a)} \;.
\ee
In this model due to its gravity nature,
the effective Newton's parameter $G_{\rm eff}$ is not any more the usual
constant $G_{N}$ but it takes the following form
(Lue, Scossimarro \& Starkman 2004)
\be
\label{Newton}
G_{\rm eff}(a)=G_{N}Q(a) \;\;\;\;\;\;\;\;Q(a)=\frac{2+4\Omega^{2}_{m}(a)}{3+3\Omega^{2}_{m}(a)} \;.
\ee

\section{The linear growth rate}
For the purpose of the present study, we first
discuss the basic equation which governs the evolution of the matter
perturbations within the framework of any
DE model (scalar or geometrical).
An important
ingredient in this analysis is the fact that
at the sub-Hubble scales the DE component
is expected to be smooth and thus one can use
perturbations only on the matter component of the
cosmic fluid (Dave, Caldwell \& Steinhardt 2002). 
In particular,
following the notations of Lue et al. (2004), 
Linder (2005), Stabenau \& Jain (2006), Uzan (2007), 
Linder \& Cahn (2007), 
Tsujikawa, Uddin \& Tavakol (2008)
and Dent, Dutta \& Perivolaropoulos (2009)
we can derive
the well known scale independent equation
of the linear matter overdensity
$\delta_{m}\equiv \delta\rho_m/\rho_m$
\be
\label{odedelta}
\ddot{\delta}_{m}+ 2H\dot{\delta}_{m}=4 \pi G_{\rm eff} \rho_{m} \delta_{m}
\ee
a solution of which is
$\delta_{m}(t) \propto D(t)$, with $D(t)$ denoting the linear growing mode
(usually scaled to unity at the present time). Notice, that
$\rho_{m}$ is the matter density.
Of course, for the scalar field DE models
[$G_{\rm eff}=G_{N}$, $Q(a)=1$], the
above equation reduces to the usual time evolution
equation for the mass density contrast (Peebles 1993), 
while in the case of modified gravity models (see Lue et al. 2004;
Linder 2004; Linder \& Cahn 2007; Tsujikawa et al. 2008; Gannouji et al. 2009)
we have $G_{\rm eff}\ne G_{N}$ (or $Q(a) \ne 1$).
Transforming equation (\ref{odedelta})
from $t$ to $a$
($\frac{d}{dt}=H\frac{d}{d\ln a}$),
we simply derive the evolution equation
of the growth factor $D(a)$
\be
\label{dela}
\frac{a^{2}}{D}\frac{d^{2}D}{da^{2}}+
\left(3+a\frac{d{\rm ln}E}{da}\right)\frac{a}{D}
\frac{dD}{da}=
\frac{3}{2}\Omega_{m}(a)Q(a) \;.
\ee
We would like to remind the reader here that
solving eq.(\ref{dela}) for
the concordance $\Lambda$ cosmology \footnote{For the usual $\Lambda$CDM
cosmological model we have
$w(a)=-1$, $\Omega_{\Lambda}(a)=1-\Omega_{m}(a)$ and $Q(a)=1$.}, we derive
the well known
perturbation growth factor (see Peebles 1993)
\be\label{eq24}
D(z)=\frac{5\Omega_{m0}
  E(z)}{2}\int^{+\infty}_{z}
\frac{(1+u)du}{E^{3}(u)} \;\;.
\ee
We would like to stress that for the $\Lambda$CDM cosmological model
we use the above equation normalized to unity at the
present time.

\begin{table*}
\caption[]{The growth data.
The correspondence of the columns is as follows: index, redshift, observed
growth rate and references.
In the final column one can find various
symbols of the data appearing in Fig.2.}
\tabcolsep 6pt
\begin{tabular}{ccccc}
\hline
Index & $z$ & $A_{obs}$& Refs.& Symbols\\ \hline \hline
1&0.17 & $0.510\pm 0.060$& Song \& Percival 2009; Percival et al. 2004& open circles\\
2&0.35 & $0.440\pm 0.050$& Song \& Percival 2009; Tegmark et al. 2006&open circles\\
3&0.77 & $0.490\pm 0.180$& Song \& Percival 2009; Guzzo et al. 2008&open circles\\
4&0.25 & $0.351\pm 0.058$& Samushia et al. 2012&open triangles\\
5&0.37 & $0.460\pm 0.038$& Samushia et al. 2012&open triangles\\
6&0.22 & $0.420\pm 0.070$& Blake et al. 2011& solid circles\\
7&0.41 & $0.450\pm 0.040$& Blake et al. 2011&solid circles\\
8&0.60 & $0.430\pm 0.040$& Blake et al. 2011&solid circles\\
9&0.78 & $0.380\pm 0.040$& Blake et al. 2011&solid circles\\
\end{tabular}
\end{table*}

\subsection{The evolution of the growth index}
As we have mentioned in the introduction,
for any type of DE,
an efficient parametrization
of the matter perturbations
is based on the growth rate of clustering originally introduced by
Peebles (1993). This is 
\be
\label{fzz221}
f(a)=\frac{d\ln D}{d\ln a}\simeq \Omega^{\gamma}_{m}(a)
\ee
which implies
\be
\label{eq244}
D(a)={\rm exp} \left[\int_{1}^{a}
\frac{\Omega_{m}^{\gamma(x)}(x)}{x} dx \right]
\ee
where $\gamma$ is the so called growth index
(see Silveira \& Waga 1994; Wang \& Steinhardt 1998; Lue et al. 2004; 
Linder 2004; Linder \& Cahn 2007; Nesseris \& Perivolaropoulos 2008).

Combining
eq.(\ref{fzz221}), eq.(\ref{dela}) and
eq.(\ref{eos22}), we find after some simple algebra
\be
\label{fzz222}
a\frac{df}{da}+f^{2}+X(a)f
= \frac{3}{2}\Omega_{m}(a)Q(a) \;,
\ee
where
\be
X(a)=\frac{1}{2}-\frac{3}{2}w(a)
\left[ 1-\Omega_{m}(a)\right] \;.
\ee
If we change variables in eq.(\ref{fzz222})
from $a$ to redshift [$\frac{d}{da}=-(1+z)^{-2}\frac{d}{dz}$]
and utilizing eqs.(\ref{domm}) (\ref{fzz221}), then
we can derive the evolution equation
of the growth index $\gamma=\gamma(z)$ [see also Polarski \& Gannouji 2008]

\ba
\label{Poll}
&& -(1+z)\gamma^{\prime}{\rm ln}(\Omega_{m})+\Omega_{m}^{\gamma}+
3w(1-\Omega_{m})(\gamma-\frac{1}{2})+\frac{1}{2}\; \nonumber \\
&& =\frac{3}{2}Q\Omega_{m}^{1-\gamma} \;,
\ea
Evaluate eq.(\ref{Poll}) at $z=0$ we have
\ba
\label{Poll1}
&& -\gamma^{\prime}(0){\rm ln}(\Omega_{m0})+\Omega_{m0}^{\gamma(0)}+
3w_{0}(1-\Omega_{m0})[\gamma(0)-\frac{1}{2}]+\frac{1}{2}\; \nonumber \\
&&=\frac{3}{2}Q_{0}\Omega_{m0}^{1-\gamma(0)}\;,
\ea
where $Q_{0}=Q(z=0)$ and $w_{0}=w(z=0)$.

It is interesting to mention here that
the last few years there have been many theoretical
speculations regarding the functional form of the growth index
and indeed various candidates have been proposed in the literature.
In this work, we decide to
phenomenologically treat the functional form of the growth index
$\gamma(z)$ as follows
\be
\gamma(z)=\gamma_{0}+\gamma_{1}y(z)\;.
\ee
In other words, the above
equation can be viewed as a first order Taylor expansion
around some cosmological quantity such as $a(z)$, $z$ and $\Omega_{m}(z)$.
Interestingly, for those $y(z)$ functions which satisfy $y(0)=0$
[or $\gamma(0)=\gamma_{0}$] one can write the
parameter $\gamma_{1}$ in terms of $\gamma_{0}$.
In this case [$\gamma^{\prime}(0)=\gamma_{1}y^{\prime}(0)$], using
eq.(\ref{Poll1}) we obtain
\be
\label{Poll2}
\gamma_{1}=\frac{\Omega_{m0}^{\gamma_{0}}+3w_{0}(\gamma_{0}-\frac{1}{2})
(1-\Omega_{m0})-\frac{3}{2}Q_{0}\Omega_{m0}^{1-\gamma_{0}}+\frac{1}{2}  }
{y^{\prime}(0)\ln  \Omega_{m0}}\;.
\ee
In brief, we present various forms of $\gamma(z)$, $\forall z$.
\begin{itemize}

\item Constant growth index (hereafter $\Gamma_{0}$ model): Here
we set $\gamma_{1}$ strictly equal to zero which implies
$\gamma=\gamma_{0}$.

\item Expansion around $z=0$ (see Polarski et al. 2008; 
hereafter $\Gamma_{1}$ model):
In this case we have $y(z)=z$. However, the latter parametrization is valid
at relatively low redshifts $0\le z \le 0.5$. In the statistical
analysis presented below we use a constant growth index namely
$\gamma=\gamma_{0}+0.5\gamma_{1}$ for $z>0.5$.

\item Interpolated parametrization
(hereafter $\Gamma_{2}$ model): Owing to the fact that the $\Gamma_{1}$ model
is valid at low redshifts we propose to use a new formula
$y(z)=z{\rm e}^{-z}$ which
connects smoothly low and high-redshifts ranges.
The latter $y(z)$ formula can be seen as a combination of $\Gamma_{1}$ model
with that of Dossett et al. (2010). Obviously, at large redshifts
$z\gg 1$ we have
$\gamma_{\infty}\simeq \gamma_{0}$.

\item Expansion around $a=1$ (Wu, Yu \& Fu 2009; Ishak \& Dosset 2009;
Belloso et al. 2011; Di Porto et al. 2012
hereafter
$\Gamma_{3}$ model): Here we use $y(z)=1-a(z)=\frac{z}{1+z}$ which implies
that for $z\gg 1$ we get
$\gamma_{\infty}\simeq \gamma_{0}+\gamma_{1}$.

\item Expansion around $\Omega_{m}=1$ (Wang \& Steinhardt 1998 
hereafter $\Gamma_{4}$ model): Now we parametrize 
$y(z)$ as follows $y(z)=1-\Omega_{m}(z)$.
For the DE models with
a constant EoS parameter $w(z)\equiv w_{0}$ one
can write $(\gamma_{0},\gamma_{1})$ only in terms of $w_{0}$
\be
\label{WA98}
\gamma_{0}=\frac{3(1-w_{0})}{5-6w_{0}} \;\;\;
\gamma_{1}=\frac{3}{125}\frac{(1-w_{0})(1-3w_{0}/2)}{(1-6w_{0}/5)^{3}} \;.
\ee
At large redshifts $\Omega_{m}(z)\simeq 1$ we get
$\gamma_{\infty}\simeq \gamma_{0}$.
Note, that the DPG cosmological model predicts
$(\gamma_{0},\gamma_{1})\simeq (\frac{11}{16},\frac{7}{256})$
(Linder 2004; Linder et al. 2007; Gong 2008).
\end{itemize}

From the above presentation it becomes evident that
for the $\Gamma_{1-3}$
parametrizations we have $y(0)=0$ and
$y^{\prime}(0)=1$, respectively. Therefore, for
the case of the $\Lambda$CDM
cosmology with $(\Omega_{m0},\gamma_{0})=(0.273,\frac{6}{11})$
eq.(\ref{Poll2}) provides $\gamma_{1}\simeq -0.0478$,
while for the case of the $\Gamma_{4}$ model we obtain
$\gamma_{1}\simeq 0.01127$ (see eq.\ref{WA98}).
In addition, based on the DGP gravity with
$(\Omega_{m0},\gamma_{0})=(0.273,\frac{11}{16})$ the $\Gamma_{1-3}$
models give $\gamma_{1}\simeq 0.05$.

\section{The Growth data}
The growth data that we will use in this work based on
2dF, SDSS and {\em WiggleZ} galaxy surveys,
for which their combination parameter of the growth rate of structure,
$f(z)$, and the redshift-dependent rms fluctuations of the linear
density field, $\sigma_8(z)$,
is available as a function of redshift, $f(z)\sigma_{8}(z)$.
The $f\sigma_{8}\equiv A$ estimator is almost a model-independent
way of expressing
the observed growth history of the universe (Song \& Percival 2009).
In particular, we will use:
\begin{itemize}
\item The 2dF (Percival et al. 2004), SDSS-LRG (Tegmark et al. 2006)
and VVDS (Guzzo et al. 2008) based growth results as collected by
Song \& Percival (2009). This sample contains 3 entries.
\item The SDSS (DR7) results (2 entries)
of Samushia et al. (2012)
based on spectroscopic data of $\sim$106000 LRGs
in the redshift bin $0.16<z<0.44$.
\item The {\it WiggleZ} results of Blake et al. (2011)
based on spectroscopic data of $\sim$152000 galaxies
in the redshift bin $0.1<z<0.9$. This dataset contains 4 entries.
\end{itemize}
In Table 1 we list the precise numerical values of the data points
with the corresponding errors bars.

\begin{figure}
\mbox{\epsfxsize=8.2cm \epsffile{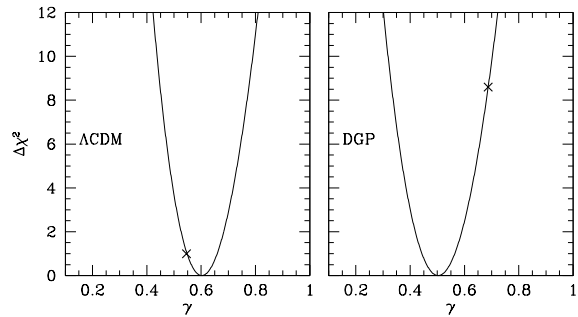}}
\caption{{\em Left Panel:}
The variance $\Delta \chi^{2}=\chi^{2}-\chi^{2}_{min}$
around the best fit $\gamma$ value for the $\Lambda$ cosmology.
Note that the cross corresponds
to $(\gamma_{\Lambda},\Delta \chi_{1\sigma}^{2})=(\frac{6}{11},1)$.
{\em Right Panel:} The statistical results in the case of
the DGP model. The corresponding cross is
$(\gamma_{DGP},\Delta \chi_{3\sigma}^{2})=(\frac{11}{16},9)$.}
\end{figure}

\section{Fitting Models to the Data}
In order to quantify the free parameters of the growth index
we perform a standard $\chi^2$ minimization procedure between $N=9$ growth
data measurements, $A_{obs}=f_{obs}(z)\sigma_{8,obs}(z)$, with the
growth values predicted by the models at the corresponding redshifts,
$A({\bf p},z)=f({\bf p},z)\sigma_{8}({\bf p},z)$ with
$\sigma_{8}({\bf p},z)=\sigma_{8,0}D({\bf p},z)$.
The vector ${\bf p}$ contains the free parameters of the model
and depending on the model. In particular,
the essential free parameters that enter in the
theoretical expectation of are:
${\bf p} \equiv (\gamma_{0},\gamma_{1},\Omega_{m0})$.
The $\chi^2$ function\footnote{Likelihoods are
normalized to their maximum values. In the present analysis we
always report $1\sigma$ uncertainties on the fitted parameters.
Note also that the total number of data points used here is
$N=9$, while the associated degrees of freedom is: {\em
  dof}$= N-k-1$, where $k$ is
the model-dependent number of fitted
parameters.
The uncertainty of the fitted parameters will be estimated, in
the case of more than one such parameters, by marginalizing one with respect
to the others.}
is defined as:
\be
\label{Likel}
\chi^{2}(z_{i}|{\bf p})=\sum_{i=1}^{N} \left[ \frac{A_{obs}(z_{i})-
A({\bf p},z_{i})}
{\sigma_{i}}\right]^{2}
\ee
where $\sigma_{i}$ is the observed growth rate uncertainty.
To this end we will use, the 
relevant to our case, 
{\em corrected} Akaike information criterion for small sample size 
(${\rm AIC}_c$; Akaike 1974, Sugiura 1978), defined,
for the case of Gaussian errors, as:
\be
{\rm AIC}_c=\chi^2_{min}+2k+\frac{2k(k-1)}{N-k-1}
\ee
where $k$ is the number of free parameters, and thus when $k=1$ then
AIC$_c=\chi_{min}^2+2$. A smaller value of AIC$_c$
indicates a better model-data fit. However, small
 differences in AIC$_c$ are not necessarily significant and therefore, in order
 to assess, the effectiveness of the different models in reproducing
 the data, one has to investigate the model pair difference 
$\Delta$AIC$_c = {\rm AIC}_{c,y} - {\rm AIC}_{c,x}$. 
The higher the value of $|\Delta{\rm AIC}_c|$, the
higher the evidence against the model with higher value of ${\rm AIC}_c$,
with a difference $|\Delta$AIC$_c| \magcir 2$ indicating a positive such evidence and
$|\Delta$AIC$_c| \magcir 6$ indicating a strong such evidence,
 while a value $\mincir 2$ indicates consistency among the two 
comparison models.
A numerical summary of the statistical analysis for
various $\gamma(z)$ parametrizations
is shown in Table 2. In general, we find that our results
are in agreement, within $1\sigma$ uncertainties, with previous
studies (Di Porto et al. 2008; Gong 2008;
Nesseris \& Perivolaropoulos; Dosset et al. 2010; Fu et al. 2009;
Basilakos 2012).

\subsection{Constant growth index}
First of all we consider the $\Gamma_{0}$ parametrization
($\gamma=\gamma_{0}$, $\gamma_{1}=0$; see section 3.1)
which implies that
the corresponding statistical vector becomes:
${\bf p} \equiv (\gamma,0,\Omega_{m0})$.
We will restrict our present analysis
to the choice $(\Omega_{m0},\sigma_{8,0})=(0.273,0.811)$
provided by WMAP7 (Komatsu et al. 2011)\footnote{For the DGP model
Gong (2008) found $\Omega_{m0}=0.278$.}.
Note that we sample $\gamma \in [0.1,1.3]$ in steps of 0.001.

\begin{figure}
\mbox{\epsfxsize=8.2cm \epsffile{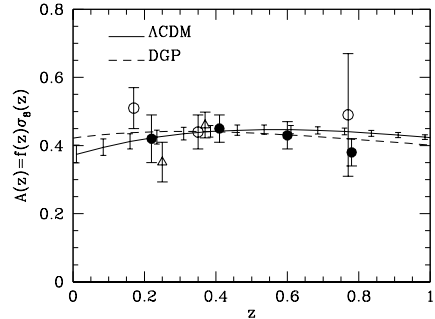}}
\caption{Comparison of the observed and
theoretical evolution of the growth
rate $A(z)=f(z)\sigma_{8}(z)$. The solid and dashed lines
correspond to $\Lambda$CDM ($\gamma=0.602$) and DGP ($\gamma=0.503$)
expansion models respectively.
The thin-line error bars 
correspond to $1\sigma$ $\gamma$-uncertainties for the $\Lambda$ cosmology. 
We do not plot the $1\sigma$ $\gamma$-uncertainties for the DGP model 
in order to avoid confusion.
The different growth datasets are
represented by different symbols (see Table 1 for definitions).}
\end{figure}

In the left panel of Fig. 1 we show
the variation of
$\Delta \chi^{2}=\chi^{2}(\gamma)-\chi^{2}_{min}(\gamma)$
around the best $\gamma$ fit value
for the concordance $\Lambda$ cosmology.
We find that the likelihood function of the growth data
peaks at $\gamma=0.602\pm 0.055$
with $\chi^{2}_{min} \simeq 7.1$
for $7$ degrees of freedom\footnote{Using eq.(\ref{eq244}) in the likelihood 
analysis for the usual $\Lambda$ cosmology we obtain $\gamma=0.595\pm 0.071$
with $\chi^{2}_{min}/dof \simeq 7.59/7$. Note that for the DGP model we only 
use eq.(\ref{eq244}).}.
Alternatively, considering the $\Lambda$CDM theoretical value of
$\gamma$ ($\equiv 6/11$) and minimizing with respect to $\Omega_{m0}$
we find $\Omega_{m0}=0.243\pm 0.034$ (see also
Nesseris \& Perivolaropoulos 2008) with $\chi^{2}_{min}/dof \simeq 7.37/7$.
Our growth index results are in
agreement within $1\sigma$ errors, to those
of Samushia et al. (2012) 
who found $\gamma=0.584\pm 0.112$. 
However,
our best-fit value is somewhat greater and almost
$1\sigma$ ($\Delta \chi_{1\sigma}^{2} \simeq 1$) away, from
the theoretically predicted value of $\gamma_{\Lambda} \simeq 6/11$
(see cross in the left panel of Fig. 1).
It is interesting to mention here that
such a small discrepancy between the theoretical
$\Lambda$CDM and observationally fitted
value of $\gamma$ has also been found
by other authors. For example, Di Porto et al. (2008)
obtained $\gamma=0.60^{+0.40}_{-0.30}$,
Gong (2008) measured $\gamma=0.64^{+0.17}_{-0.15}$
while Nesseris \& Perivolaropoulos (2008)
found $\gamma=0.67^{+0.20}_{-0.17}$. Recently,
Basilakos (2012) and Hudson \& Turnbull (2012) 
using a similar analysis found 
$\gamma=0.613^{+0.088}_{-0.083}$ and $\gamma=0.619\pm 0.054$ 
respectively. 

Concerning the DGP model (see the right panel of Fig. 1)
the best fit parameter is $\gamma=0.503 \pm 0.06$
with $\chi^{2}_{min}/dof \simeq 5.32/7$.
If we fix the value of
$\gamma$($\equiv 11/16$) to that predicted by
the DGP model we find a rather large value
of the dimensionless matter density at the present time,
$\Omega_{m0}=0.380\pm 0.042$
with $\chi^{2}_{min}/dof \simeq 5.38/7$.

The value of AIC$_{C,DGP}$($\simeq 7.32$) is smaller than the
corresponding $\Lambda$CDM value which indicates 
that the DGP model ($\gamma_{DGP}=0.503$) appears now to fit
slightly better than the usual $\Lambda$ cosmology the growth data.
However, the small $|\Delta$AIC$_c|$ value (ie., $\sim 1.8$)
indicates that the two comparison models represent the growth data at a
statistically equivalent level.
On the other hand form the right panel of Fig.1, it becomes clear that
the best-fit $\gamma$ value is much lower and almost
$3 \sigma$ ($\Delta \chi_{3\sigma}^{2}\simeq 9$) away, from
$\gamma_{DGP} \simeq 11/16$
(see cross in the right panel of Fig. 1) implying that
the growth data disfavor the DGP gravity.
We would like to stress here that
the above observational DGP constraints
are in excellent agreement
with previous studies. Indeed, Wei (2008) found 
$\gamma =0.438^{+0.126}_{-0.111}$. Also
Gong (2008) and Dosset et al. (2010)
obtained $\gamma =0.55^{+0.14}_{-0.13}$
and $\gamma =0.483^{+0.113}_{-0.088}$ respectively.
In Fig. 2, we plot the measured $A_{obs}(z)$
with the estimated growth rate
function, $A(z)=f(z)\sigma_{8}(z)$
[see $\Lambda$CDM - solid line and DGP - dashed line].


The goal from the above discussion
is to give the reader the
opportunity to appreciate the relative strength and precision
of the different methods used in order to constrain the growth index.
It becomes evident that
with the combined high-precision $f\sigma_{8}$
growth rate data of
Song \& Percival (2009), Samushia et al. (2012)
and Blake et al. (2011)
we have achieved to place
quite stringent constraints on $\gamma$.

\begin{table*}
\caption[]{Statistical results for the combined growth data
(see Table 1): The $1^{st}$ column
indicates the expansion model, the $2^{nd}$ column corresponds to
$\gamma(z)$ parametrizations
 appearing in section 3.1.
$3^{rd}$ and $4^{rth}$ columns show the $\gamma_{0}$ and $\gamma_1$ best values.
The remaining columns present the goodness-of-fit statistics 
(reduced $\chi^{2}$ and AIC$_{c}$).}
\tabcolsep 6pt
\begin{tabular}{cccccc}
\hline
Expansion Model & Parametrization Model & $\gamma_{0}$& $\gamma_{1}$& $\chi_{min}^{2}/dof$ &${\rm AIC}_{c}$\\ \hline \hline
$\Lambda$CDM&$\Gamma_{0}$&$0.602\pm 0.055$& 0 &7.10/7&9.10  \\
            &$\Gamma_{1}$&$0.400^{+0.086}_{-0.080}$& $0.603\pm 0.241$ &5.74/6&10.41 \\
            &$\Gamma_{2}$&$0.311^{+0.085}_{-0.080}$& $1.221\pm 0.343$ &5.26/6&9.94 \\
            &$\Gamma_{3}$&$0.345^{+0.085}_{-0.080}$& $1.006\pm 0.314$ &5.06/6 &9.74\\
DGP         &$\Gamma_{0}$&$0.503\pm 0.060$& 0 &5.32/7& 7.32  \\
            &$\Gamma_{1}$&$0.441^{+0.094}_{-0.090}$& $0.164\pm 0.221$ &5.10/6&9.73 \\
            &$\Gamma_{2}$&$0.401^{+0.094}_{-0.090}$& $0.384\pm 0.320$ &5.00/6 &9.66\\
            &$\Gamma_{3}$&$0.412^{+0.093}_{-0.090}$& $0.321\pm 0.290$ &4.94/6&9.60 \\

\end{tabular}
\end{table*}

\subsection{The $\Gamma_{1-4}$ parametrizations}
After we have presented the simplest version of the growth index,
it seems appropriate to discuss the observational constraints
on the time varying growth index, $\gamma(z)$.
Following the considerations exposed in section 3.1, hereafter we
will set 
${\bf p}=(\gamma_{0},\gamma_{1},0.273)$
in equation (\ref{Likel}).
In Fig. 3 ($\Lambda$CDM model) and Fig. 4 (DGP model)
we present the
results of our statistical analysis
for the $\Gamma_{1}$ (upper left panel),
$\Gamma_{2}$ (upper right panel),
$\Gamma_{3}$ (bottom left panel) and $\Gamma_{4}$ (bottom right panel)
parametrizations in the $(\gamma_{0},\gamma_{1})$ plane in which
the corresponding contours are
plotted for 1$\sigma$, 2$\sigma$ and 3$\sigma$ confidence levels.
Notice, that we sample $\gamma_{0} \in [0.1,1.3]$ and
$\gamma_{1} \in [-2.2,2.2]$ in steps of 0.001.
The theoretical $(\gamma_{0},\gamma_{1})$ values (see section 3.1)
in the $\Lambda$CDM and DGP expansion models indicated by the
crosses. Overall, we find that the predicted $\Lambda$CDM
$(\gamma_{0},\gamma_{1})$
solutions of the $\Gamma_{1-4}$ parametrizations
remain close to the $1\sigma$ borders ($\Delta \chi_{1\sigma}^{2}\simeq 2.30$;
see crosses in Fig.3).
Regarding the DGP model (see Fig.4) we
would like to stress that the predicted $(\gamma_{0},\gamma_{1})$ values
approach the $3 \sigma$ borders
($\Delta \chi_{3\sigma}^{2}\simeq 11.83$; see crosses on Fig.4)
of the $\gamma_{0}-\gamma_{1}$ contours. Obviously,
this is a clear indication 
that the current growth data can not accommodate
the DGP gravity model.

\begin{figure}
\mbox{\epsfxsize=8.2cm \epsffile{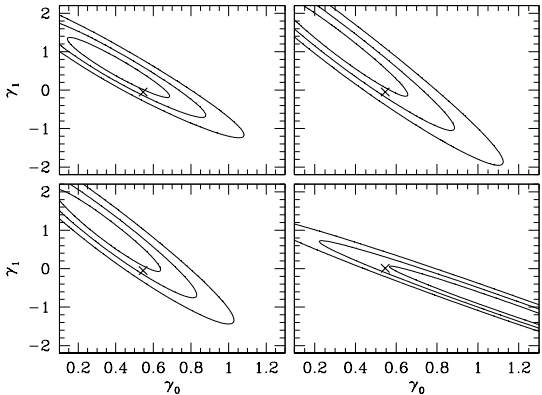}}
\caption{The $\Lambda$CDM {\em expansion model:} Likelihood contours (for
$\Delta \chi^{2}=-2{\rm ln}{\cal L}/{\cal L}_{\rm max}$ equal to 2.30, 6.18
and 11.83, corresponding
to 1$\sigma$, 2$\sigma$ and $3\sigma$ confidence levels) in the
$(\gamma_{0},\gamma_{1})$. The upper left and right panels show the
results based on the $\Gamma_{1-2}$ parametrizations.
In the bottom left and right panels we present the contours of the
$\Gamma_{3-4}$ parametrizations (for more details see section 3.1).
We also include the theoretical $\Lambda$CDM
$(\gamma_{0},\gamma_{1})$ values given in section 3.1.}
\end{figure}

\begin{figure}
\mbox{\epsfxsize=8.2cm \epsffile{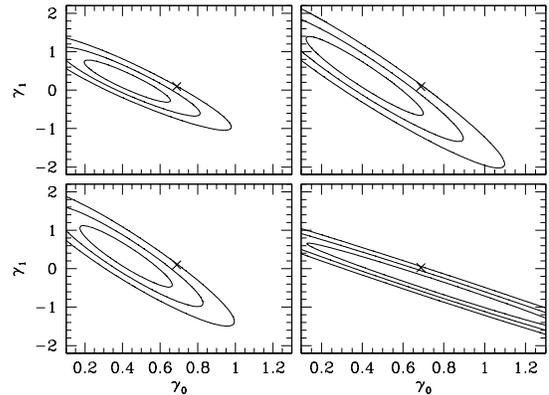}}
\caption{The Likelihood contours for the DGP expansion model
(for more definitions see caption of figure 3).
Here the crosses correspond to the theoretical DGP
$(\gamma_{0},\gamma_{1})$ pair
provided in section 3.1.}
\end{figure}

Below we briefly discuss the main statistical results:
\noindent
(a) $\Gamma_{1}$ parametrization:
For the usual $\Lambda$ cosmology the likelihood function
peaks at $\gamma_{0}=0.40^{+0.086}_{-0.080}$ and
$\gamma_{1}=0.603\pm 0.241$
with $\chi_{min}^{2}/dof \simeq 5.74/6$, while in the case of the DGP gravity
we obtain $\gamma_{0}=0.441^{+0.094}_{-0.090}$ and
$\gamma_{1}=0.164\pm 0.221$
with $\chi_{min}^{2}/dof \simeq 5.10/6$.
Interestingly, the use of the combined growth data
provides a significant improvement
in the derived $(\gamma_{0},\gamma_{1})$ constraints with respect
to the previous studies (Di Porto et al. 2008; Gong 2008;
Nesseris \& Perivolaropoulos 2008; Dosset et al. 2010; Fu et al. 2009).

Such an improvement is to be expected because
the {\em WiggleZ} and the SDSS-DR7
surveys measure $f(z)\sigma_{8}(z)$ to
within $8-17\%$ (Blake et al. 2011; Samushia et al. 2012)
in every redshift bin,
in contrast to the old growth rate data (Song \& Percival 2009)
in which the corresponding accuracy lies in the interval $12-37\%$.

\noindent
(b) Now we concentrate on the $\Gamma_{2}$ and $\Gamma_{3}$:
parametrizations:
We find that within $1\sigma$
errors we can put some constraints on the free parameters. In particular,
the best fit values are: (i) $\Lambda$CDM: for $\Gamma_{2}$ we have
$\gamma_{0}=0.311^{+0.085}_{-0.080}$,
$\gamma_{1}=1.221\pm 0.343$ ($\chi_{min}^{2}/dof \simeq 5.26/6$)
while for $\Gamma_{3}$ we get
$\gamma_{0}=0.345^{+0.085}_{-0.080}$,
$\gamma_{1}=1.006\pm 0.314$ ($\chi_{min}^{2}/dof \simeq 5.06/6$)
and (ii) DGP: in the case of
$\Gamma_{2}$ model we obtain
$\gamma_{0}=0.401^{+0.094}_{-0.090}$,
$\gamma_{1}=0.384\pm 0.320$ ($\chi_{min}^{2}/dof \simeq 5.00/6$) and for
$\Gamma_{3}$ we find $\gamma_{0}=0.412^{+0.093}_{-0.090}$,
$\gamma_{1}=0.321\pm 0.290$. In the latter case 
the reduced $\chi_{min}^{2}$ is $\simeq 4.94/6$.

\noindent
(c) $\Gamma_{4}$ parametrization: In this case
the $\gamma_{0}$ is strongly degenerate with $\gamma_{1}$
(see bottom left panels in Figs.3,4). Indeed,
we can provide the following $\gamma_{1}-\gamma_{0}$ relations:

\begin{equation}
\gamma_{1}=\left\{ \begin{array}{cc}
       1.134(\pm 0.005)-1.879(\pm 0.006)\gamma_{0}   &
       \mbox{$\Lambda$CDM}  \\
       0.887(\pm 0.005)-1.749(\pm 0.006)\gamma_{0} & \mbox{DGP}
       \end{array}
        \right.
\end{equation}

\begin{figure}
\mbox{\epsfxsize=8cm \epsffile{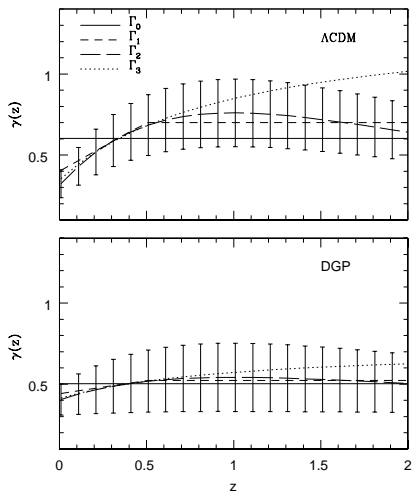}}
\caption{{\it Upper Panel:} 
The evolution of the growth index
for the $\Lambda$CDM model.
The lines correspond to $\Gamma_{0}$ (solid), $\Gamma_{1}$ (short dashed),
$\Gamma_{2}$ (long dashed) and 
$\Gamma_{3}$ (dot-dashed).
The error bars 
correspond to $1\sigma$ $\gamma$-uncertainties for the $\Gamma_{2}$ 
parametrization.
We do not plot the $1\sigma$ $\gamma$-uncertainties for the other 
parametrizations in order to avoid confusion.
{\it Bottom Panel:} 
The evolution of the growth index
for the DGP cosmological model.}
\end{figure}

Finally, as we have already mentioned in Table 2, one may see a
more compact presentation of our statistical results.
For both cosmological ($\Lambda$CDM and DGP) models the 
information theory pair model characterization parameter, $\Delta$AIC$_c$,
indicates that all the $\gamma(z)$ functional forms explored in 
this study are statistically equivalent in
representing the growth rate data, since $|\Delta$AIC$_c|<2$ for any
pametrization pair. 
In Fig. 5 we present the evolution of the growth index
for various parametrizations. 
In the case of the concordance $\Lambda$ cosmology (upper panel of fig.5) 
the relative growth index difference of the various fitted $\gamma(z)$ 
models indicates that 
the $\Gamma_{1-3}$ models have a very similar
redshift dependence for $z \le 0.5$, while the $\Gamma_{3}$ parametrization
shows very large such deviations 
for $z>0.5$. Based on the DGP gravity model (bottom panel of fig.5) 
we observe that the $\Gamma_{1-2}$ parametrizations provide
a similar evolution of the growth index. The $\Gamma_{3}$ parametrization
shows large deviations at large redshifts $z\ge 1.5$. 
However the large $\gamma(z)$ errors appear in fig.5 are due to
the large uncertainty of the $\gamma_{1}$ 
fitted parameter, implying that more and accurate
data are essential in order to distinguish among the 
different $\gamma(z)$ functional forms.

\section{Conclusions}
It is well known that the so called growth index $\gamma$ plays a key role in
cosmological studies because it can be used as a useful tool in order
to test Einstein's general relativity on cosmological scales.
We have utilized the recent growth rate data
provided by the 2dFGRS, SDSS-LRG, VVDS
and {\em WiggleZ} galaxy surveys,
in order to constrain the growth index.
Performing a likelihood analysis for various
$\gamma(z)$ parametrizations, we argue that the
use of the above combined growth data
places the most stringent constraints on the value of the
growth index. 
Overall, considering a $\Lambda$CDM expansion model
we find that the observed growth index is in agreement, within 
$1\sigma$ errors, with the theoretically 
predicted value of $\gamma_{\Lambda}\simeq 6/11$.
In contrast, for the DGP expansion model we find that the measured growth index
is almost $3 \sigma$ away from the corresponding theoretical value
$\gamma_{DGP} \simeq 11/16$ which implies that the present growth data
can not accomodate the DGP gravity model.
Finally, considering a time varying growth index parametrization
namely $\gamma(z)=\gamma_{0}+\gamma_{1}y(z)$ 
[where $y(z)=z,ze^{-z},1-a(z)$ and $1-\Omega_{m}(z)$] 
we find that although the $\gamma_{0}$ parameter is tightly
constrained, the $\gamma_{1}$ parameter remains 
weakly constrained. Hopefully, with the next 
generation of surveys, based on {\em Euclid} and {\em BigBOSS},
we will be able to put strong constraints on $\gamma_{1}$
and thus to check departures from $\gamma=const.$


\end{document}